\documentclass[]{aa}
\usepackage{graphicx}
\usepackage{color}
\usepackage{epstopdf}
\usepackage{txfonts}
\usepackage{hyperref}

\usepackage{fixltx2e}
\usepackage{dblfloatfix}
\usepackage{float}

\usepackage[round]{natbib}
\begin{document}

\title{Statistical properties of coronal hole rotation rates: Are they linked to the solar interior?}

\author{S. R. Bagashvili\inst{1,2,4}, B.M. Shergelashvili\inst{3,2,4}, D. R. Japaridze\inst{2}, B.B. Chargeishvili\inst{2}, A. G. Kosovichev\inst{5}, V. Kukhianidze\inst{2}, G. Ramishvili\inst{2}, T.V. Zaqarashvili\inst{3,2,6} , S.  Poedts\inst{1},  M. L. Khodachenko\inst{3}, P. De Causmaecker\inst{4}
}
\institute{Center for Mathematical Plasma Astrophysics, Department of Mathematics, KU Leuven, 200 B, B-3001, Leuven, Belgium \\
                            \and
            Abastumani Astrophysical Observatory at Ilia State University, University St. 2, Tbilisi, Georgia\\
                            \and
            Space Research Institute, Austrian Academy of Sciences, Schmiedlstrasse 6, 8042 Graz, Austria \\
                            \and
            Combinatorial Optimization and Decision Support, KU Leuven campus Kortrijk, E. Sabbelaan 53, 8500 Kortrijk, Belgium \\
                            \and
            New Jersey Institute of Technology, Newark, NJ 07103, USA\\
                            \and
            Institute of Physics, IGAM, University of Graz, Universit\"atsplatz 5, 8010 Graz, Austria \\
}

%\date{Received 30 December 20016 / Accepted 25 May 2017}

\abstract{The present paper discusses results of a statistical study of the characteristics of coronal hole (CH) rotation in order to find connections to the internal rotation of the Sun.}{The goal is to measure CH rotation rates and study their distribution over latitude and their area sizes. In addition, the CH rotation rates are compared with the solar photospheric and inner layer rotational profiles.}{We study coronal holes observed within $\pm60^{\circ}$ latitude and longitude from the solar disc centre during the time span from the 1 January 2013 to 20 April 2015, which includes the extended peak of solar cycle 24. We used data created by the Spatial Possibilistic Clustering Algorithm (SPoCA), which provides the exact location and characterisation of solar coronal holes using $SDO/AIA\;193\;{\AA}$ channel images. The CH rotation rates are measured with four-hour cadence data to track variable positions of the CH geometric centre. }{North-south asymmetry was found in the distribution of coronal holes: about 60 percent were observed in the northern hemisphere and 40 percent were observed in the southern hemisphere. The smallest and largest CHs were present only at high latitudes. The average sidereal rotation rate for 540 examined CHs is $13.86\;(\pm0.05)\;^{\circ}/d$. }{The latitudinal characteristics of CH rotation do not match any known photospheric rotation profile. The CH angular velocities exceed the photospheric angular velocities at latitudes higher than 35-40 degrees. According to our results, the CH rotation profile perfectly coincides with tachocline and the lower layers of convection zone at around $0.71\;R_{\sun}$; this indicates that CHs may be linked to the solar global magnetic field, which originates in the tachocline region.}


\titlerunning{Statistical properties of coronal hole rotation rates}

\authorrunning{Bagashvili et al.}

\maketitle

%=====================
\section{Introduction}
%=====================

The investigation of rotation profiles of different layers of the solar interior and its atmosphere is important for our understanding of the physical processes driving the solar activity and the connections between the  Sun and heliosphere. It is one of the goals of the forthcoming Solar Orbiter mission and other observational missions as well. In fact, studies of the solar rotation rates already began in the 17th century after the first discovery of sunspots. Despite a large amount of accumulated data, we still need more observations and improved mathematical modelling to get more insight into the interconnection of the various physical processes operating at different radial distances inside and outside the Sun.

An important dynamical indicator that is connected to both the solar interior and its atmosphere is the solar wind. The solar wind, a global plasma flow from the Sun into the interplanetary space, is characterised by a variety of physical properties. Regarding characteristic flow patterns, the solar wind is bimodal and consists of the so-called fast and slow winds, and includes co-rotating interaction regions (CIRs) that represent a transitional flow pattern with a significant cross gradient of the speed. The characteristic velocities, location, morphology, and other plasma parameters of this flow pattern vary with the solar cycle \citep{Miralles2001, Miralles2002, Miralles2004}. For instance, the speed of the fast solar wind is about $600-800\;km/s$ along 'open' magnetic field lines that are identified as coronal holes \citep{Krieger1973, Pneuman1973, Noci1973}. Coronal holes (CHs) are large, magnetically unipolar \citep{Wilcox1968, Altschuler1972, Timothy1975} low-density plasma \citep{Munro1972} regions in the solar corona. In epochs of the solar minimum, large CHs are mainly concentrated at high latitudes around the polar regions, while the main streamer belt and the heliospheric current sheet are located near
the equator, enabling the formation of the slow solar wind flow pattern with approximately $400\;km/s$ velocity. During solar maxima, such a strict dichotomy of the fast and slow wind patterns disappears as the solar atmosphere becomes populated with many small fragmentised CHs and streamers. This situation results in a more or less homogenous mixture of the fast, slow, and CIR flow patterns both in latitude and longitude. Therefore, the study of the rotational and statistical properties of CHs (as containers of fast solar wind patterns) is very important. As a matter of fact, their distributions are of a large-scale nature and seemingly are linked to the global solar magnetic field. Consequently, they are often considered as indicators of the large-scale magnetic field in the solar corona \citep{Obridko1989, Bumba1995}. Recently, \citet{Bilenko2016} investigated some temporal and spatial regularities in CH distributions during several solar cycles. Their results indicate that the longitudinal distribution of non-polar CHs follows all structural changes of the global magnetic field. In general, the formation and acceleration mechanisms of various flow patterns are linked to the understanding of the near Sun physical processes and their understanding requires investigation of the heliosphere as a global system  \citep[for instance see recently reported analytical modelling;][]{Shergelashvili2012}. Currently, the attention is stimulated by the forthcoming Solar Orbiter space mission.

The first attempts to study rotation rate distributions of coronal holes over different latitudes indicated rigid rotational profiles \citep{Wagner1975, Wagner1976, Timothy1975, Adams1976, Bohlin1977, Hiremath13, Japaridze2015}. However, more recent studies showed the presence of pronounced differential rotations \citep{Shelke1985, Obridko1988, Insley1995, Navarro-Peralta1994}. Despite several improvements of the observational methods and the automation of search algorithms, final conclusions regarding this issue are currently not available. Some researchers report that the CH rotation rates vary with the solar cycle \citep{Harvey1987, Obridko1988, Navarro-Peralta1994}. During solar minimum CHs rotate more rigidly unlike to the maximum. In this regard \citet{Nash1988} concluded that this difference may be a consequence of connection between CHs and randomly emerging photospheric magnetic fields.  In addition, \citet{Mancuso2011} reported that coronal magnetic structures rotate less differentially and much faster than the lower layers of the solar atmosphere, especially above of $40^{\circ}$ latitude.

In the current paper, we investigate the CH rotational dynamics covering the period from 1 January 2013 to 20 April 2015, which included the prolonged peak of solar cycle 24. The results are analysed regarding the distributions over latitudinal locations and effective areas of CHs separately. Finally, we compare the CH rotational rates to the photospheric rotation and rotational profiles of the solar interior obtained from helioseismic measurements \citep{Schou1998}.

%=====================
\section{Data and method}\label{secobservations}
%=====================

%=====================
\subsection{Observational data}
%=====================

We have retrieved data of the Atmospheric Imaging Assembly (AIA) on board the Solar Dynamics Observatory (SDO) from the Heliophysics Events Knowledgebase (HEK) via the SolarSoft (SSW) IDL tools for the period from 1 January 2013 to 20 April 2015, which corresponds to the maximum of solar cycle 24. Using the database downloaded from HEK, we created catalogues of coronal holes from 2010 up to date. This catalogue is created in the framework of the SOLSPANET project ("Solar and Space Weather Network of Excellence") and is publicly available at \url{http://www.solspanet.eu}.

The AIA is equipped with four $4096\times4096$ detectors with a pixel size of 0.6 arcsec. The AIA records a full set of near-simultaneous images with a 12 sec cadence. The instrument contains 10 different wavelength channels; 7 extreme ultraviolet (EUV) channels (131, 171, 193, 211, 335, 94, and $304\;{\AA}$), which produce the full-disc images of the corona and transition region; and 3 channels in white light and UV bands \citep{Lemen12, Boerner11}.

Even though CHs are clearly visible in X-ray and EUV images as dark cavities, it is difficult to detect these objects using a single EUV wavelength. Filaments also resemble dark structures in EUV images, and in some cases, it is hard to distinguish and identify CHs. During the first studies of CHs, they were identified manually by observers. In past years, various automatic detection methods have been developed, most of which are based on the intensity threshold method \citep{Delouille2007, Kirk2009, Krista2009, Rotter2012}. Recently, geometrical classification \citep{Reiss2014} and supervised machine learning methods \citep{Reiss2015} of these coronal features have been developed, which represent more advanced CH detection approaches. In this work, we used data from the Spatial Possibilistic Clustering Algorithm (SPoCA), which is based on fuzzy clustering of intensity values \citep{Verbeeck13, Verbeeck2014}. This algorithm provides the exact localisation and characterisation of solar active regions and coronal holes on $SDO/AIA\;193\;{\AA}$ wavelength images. This channel provides images of the corona at approximately $3700\pm700\;km$ above the photosphere \citep{Aschwanden13}. The SPoCA yields CHs data every four hours and gives us the possibility to track the motion of individual CHs as described in \citet{Verbeeck13}.

%=====================
\subsection{Method of analysis}
%=====================

The rotational velocities of particular CHs were calculated using the position of the geometric centre of a polygon for each object. This method is justified because we measure the velocities of CHs every four hours. This time interval is too short for significant modifications of the shape and area of the CHs. Therefore, the velocity of the geometric centre of the polygon corresponds to the velocity of the CH itself.

Obviously, there are exceptions, in particular, significant changes might occur in the surface area or shape of the CHs, and then the velocity of the geometric centre may no longer reflect the movement of the CH. The CHs might also be broken in a few parts or merge with other CHs. Such cases are excluded from our statistical analysis.

To investigate the probable differential motion of coronal holes, we divided the solar surface into 32 latitudinal zones, each 3.75 degrees wide with the following intervals: from 0 to $\pm3.75^{\circ}$, $\pm3.75-7.5^{\circ}$, $\pm7.5-11.25^{\circ}$, ... , $\pm56.25-60^{\circ}$. We then classified the CHs into three groups, viz.\ relatively small-, medium-, and large-sized objects, namely with an area less than $10\;000\;Mm^{2}$, between $10\;000-40\;000\;Mm^{2}$ and larger than $40\;000\;Mm^{2}$, respectively. The method of CH area estimation  is standard \citep{Verbeeck2014} and the values are given in related catalogues (i.e. SPoCA and SOLSPANET).

%%%%%%%%%%%%%%%%%%%%%%%%%%%%%%%%%%%%%%%%%%%%%%%%%%%%%%%%%%%%%%%%%%%%%%%%%%%%%%%%%%%%%%
\begin{figure}[h]
\vspace*{1mm}
\begin{center}
\includegraphics[scale=1.0]{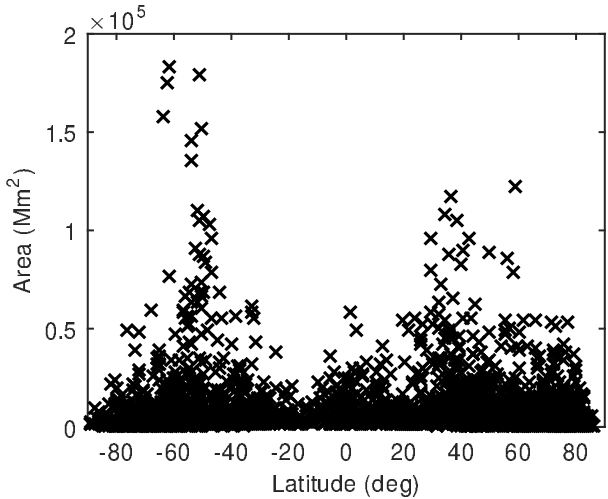}
\end{center}

\caption{Distribution of 3056 CHs with latitude and areas during the years 2013-2015. Black crosses represent the individual CHs.} \label{Fig1}
\end{figure}
%%%%%%%%%%%%%%%%%%%%%%%%%%%%%%%%%%%%%%%%%%%%%%%%%%%%%%%%%%%%%%%%%%%%%%%%%%%%%%%%%%%%%%

%%%%%%%%%%%%%%%%%%%%%%%%%%%%%%%%%%%%%%%%%%%%%%%%%%%%%%%%%%%%%%%%%%%%%%%%%%%%%%%%%%%%%%
\begin{figure*}[h]
\vspace*{1mm}
\begin{center}
\includegraphics[scale=1.0]{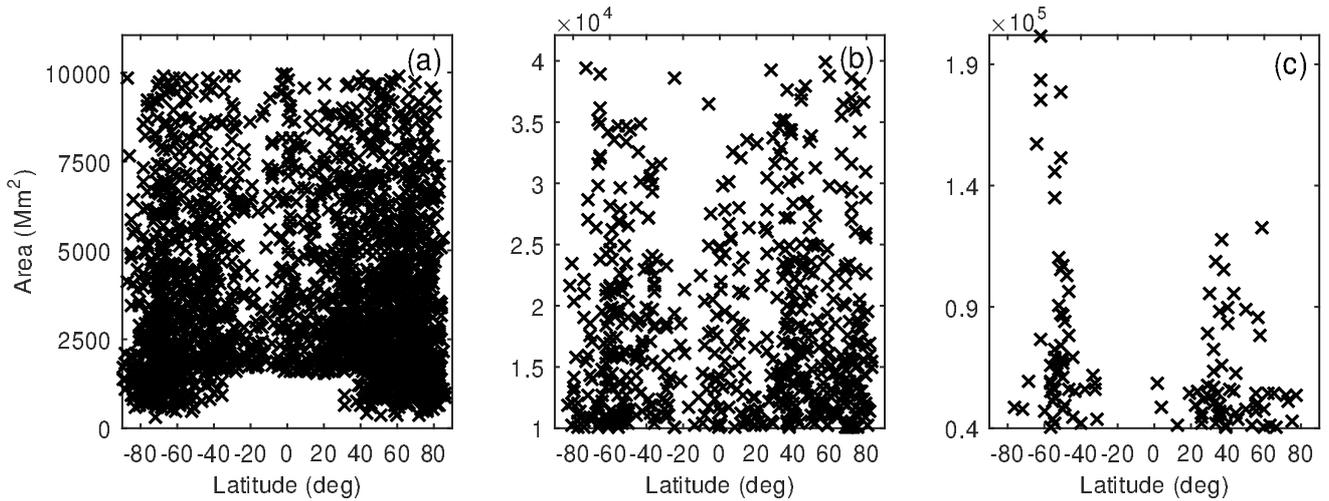}
\end{center}

\caption{Distribution of CHs in years 2013-2015 by area groups. Black crosses represent individual CHs. Panel (a) represents CHs with an average area less than $10\;000\;Mm^{2}$; panel (b) shows CHs between $10\;000-40\;000 \;Mm^{2}$; and panel (c) represents CHs larger than $40\;000\;Mm^{2}$.}
\label{Fig2}
\end{figure*}
%%%%%%%%%%%%%%%%%%%%%%%%%%%%%%%%%%%%%%%%%%%%%%%%%%%%%%%%%%%%%%%%%%%%%%%%%%%%%%%%%%%%%%

According to these criteria, we developed a special programme using the Interactive Data Language (IDL). This code uses the heliographic coordinates of the CHs geometric centres to calculate the displacement of each object after every four hours. Finally, it provides a measurement of the angular velocities in the different latitudinal zones and for the three area groups. Using this method, we calculated in total angular velocities of 540 objects during the studied time span.

%=====================
\section{Results}\label{secresults}
%=====================

%=====================
\subsection{Distribution of coronal holes over latitude and areas}
%=====================

We studied the CH distributions with respect to latitudes and CH areas. The results are presented in Fig~\ref{Fig1} and Fig~\ref{Fig2}. The horizontal axis corresponds to the latitude; the vertical axis denotes the average area of the CHs. Black crosses correspond to individual CHs, for each graph. Fig~\ref{Fig1} shows the 2D distribution for all detected CHs. Fig~\ref{Fig2} shows the distribution separately for small (panel (a)), medium (panel (b)), and large (panel (c)) CHs.

As we can see, in this epoch (solar maximum) CHs are observed at all latitudes. This is in contrast to solar minimum \citep{toma11}, when they are only found near high latitudes and at the poles. The prime characteristic of this distribution is that small CHs, with an area less than $2000\;Mm^{2}$,  are not observed in the equatorial zone, as is clear from Fig~\ref{Fig2}, panel (a). According to panel (b) of Fig~\ref{Fig2}, extremely large CHs are not found is this region either.

We noticed a north-south asymmetry in the distribution of coronal holes during the studied time window. In the northern hemisphere, coronal holes were observed about 1.48 times more frequently (1822, $59,62\;\%$) than in the southern hemisphere (1234, $40,38\;\%$). Accordingly, the total number of detected CHs was 3056.

In general, the sizes of CHs vary within a quite wide range: the maximum size can be one-third or more of the solar disc area. In our case, the smallest detected coronal hole was 71$\;Mm^{2}$ ($0.0012\;\%$ of the solar disc surface area). The largest coronal hole occupied about $203.257\;Mm^{2}$ ($9.24\;\%$).

\begin{figure}[h]
\vspace*{1mm}
\begin{center}
\includegraphics[scale=1.0]{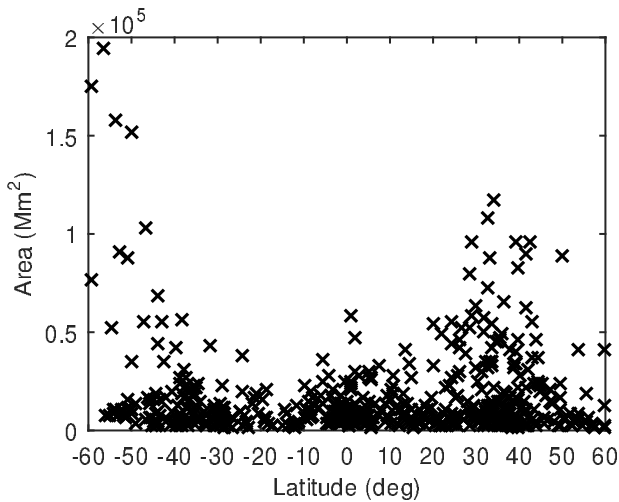}
\end{center}

\caption{Distribution of the 540 selected CHs over latitude and area. Black crosses represent the individual CHs.}
\label{Fig3}
\end{figure}

Finally, as mentioned above, for the study of the distribution of CHs over the solar disc we had registered in total 3056 CHs. However, we selected only 540 CHs for the calculation of the rotation rates, using the following criteria: (i)~They must be located on the disc within the 120$^{\circ}$ latitude and 120$^{\circ}$ longitude box; (ii)~the lifetime must be at least 12 hours (i.e.\ they must appear in at least three snapshots); and (iii)~the form and the area of the CH polygon must not change significantly during the observational time. The latitudinal distribution of these selected objects is not homogenous (Fig~\ref{Fig3}) and is similar to the distribution of the whole CH set shown in Fig.~1. Although we consider $3.75^{\circ}$ wide latitudinal zones, at least four CHs were studied in each zone, while less than 10  objects were present in only six latitudinal zones.

%%%%%%%%%%%%%%%%%%%%%%%%%%%%%%%%%%%%%%%%%%%%%%%%%%%%%%%%%%%%%%%%%%%%%%%%%%%%%%%%%%%%%%
%%%%%%%%%%%%%%%%%%%%%%%%%%%%%%%%%%%%%%%%%%%%%%%%%%%%%%%%%%%%%%%%%%%%%%%%%%%%%%%%%%%%%%

%%%%%%%%%%%%%%%%%%%%%%%%%%%%%%%%%%%%%%%%%%%%%%%%%%%%%%%%%%%%%%%%%%%%%%%%%%%%%%%%%%%%%%
\begin{figure*}[h]
\vspace*{1mm}
\begin{center}
\includegraphics[scale=1.0]{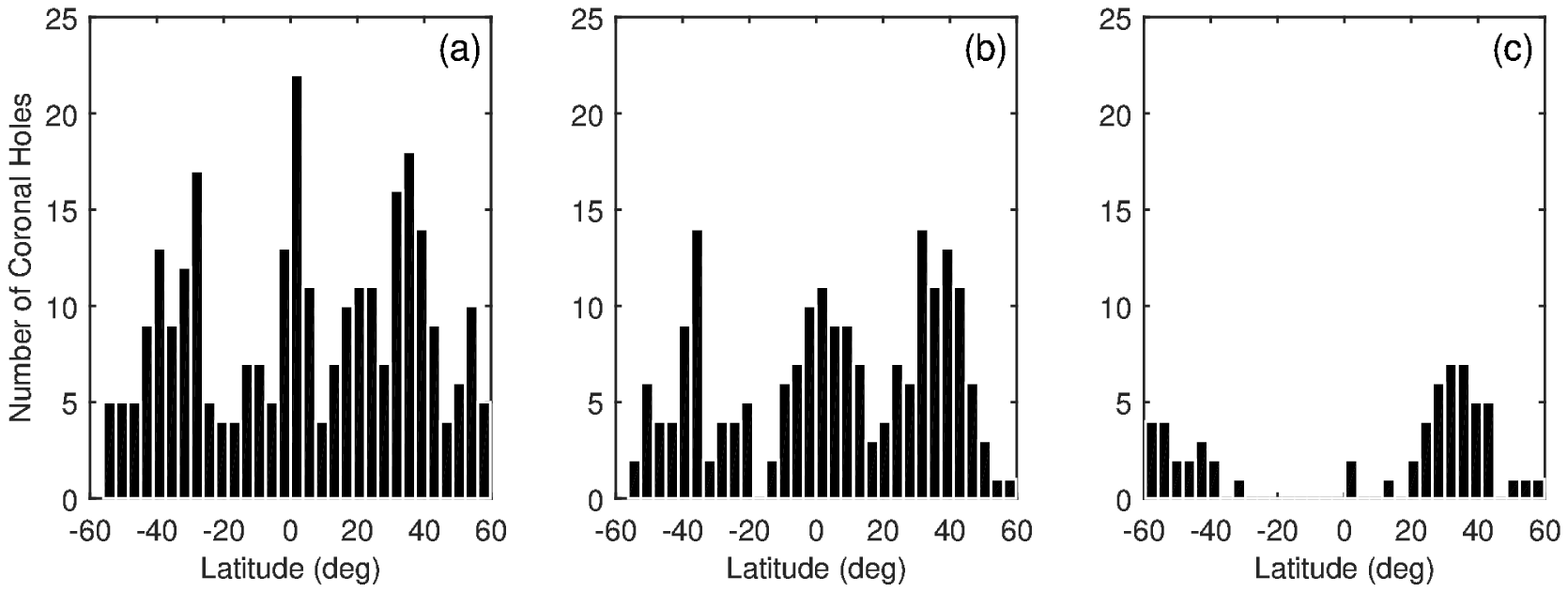}
\end{center}

\caption{Distribution of investigated CHs by latitudinal zones. Panel (a) represents coronal holes with average area less than $10\;000\;Mm^{2}$; panel (b) shows CHs with area in the range $10\;000-40\;000\;Mm^{2}$; and panel (c) shows CHs larger than $40\;000\;Mm^{2}$.}
\label{Fig4}
\end{figure*}
%%%%%%%%%%%%%%%%%%%%%%%%%%%%%%%%%%%%%%%%%%%%%%%%%%%%%%%%%%%%%%%%%%%%%%%%%%%%%%%%%%%%%%

The latitudinal distribution of CHs divided by the area groups is presented in Fig~\ref{Fig4}. The distribution of CHs having areas less than $10\;000\;Mm^{2}$ is similar to the distribution of all CHs because they constitute the majority of the observed objects. As a matter of fact, approximately 53$\%$ of the total number of CHs is in this category (i.e.\ 285 CHs, Fig~\ref{Fig4} panel (a)). Also, 195 average-sized objects were studied in all latitudinal zones (Fig~\ref{Fig4} panel (b)). Coronol holes with areas larger than $40\;000\;Mm^{2}$ are the rarest; only 60 objects fall in this category (Fig~\ref{Fig4} panel (c)).

\subsection{Rotation rates of coronal holes}

%%%%%%%%%%%%%%%%%%%%%%%%%%%%%%%%%%%%%%%%%%%%%%%%%%%%%%%%%%%%%%%%%%%%%%%%%%%%%%%%%%%%%%%
\begin{table*}    %[h]
\caption{Values of A, B, and C coefficients for the sidereal rotation of CHs after fitting the data on the pattern $A+Bsin^{2}(\theta)+Csin^{4}(\theta)$. All coefficient are measured in ${^\circ}/d$.}
\centering
    \begin{tabular}{ | l | p{1.6cm} | p{1.6cm} | p{1.6cm} | p{1.6cm} | p{1.95cm} | p{1.95cm} | p{1.6cm} | p{1.6cm} | }
    \hline\hline
   Coeff.   & South hemisphere (all CHs) & North hemisphere (all CHs) & South hemisphere (small CHs) & North hemisphere (small CHs) & South hemisphere (medium CHs) & North hemisphere (medium CHs) & South hemisphere (large CHs) & North hemisphere (large CHs) \\ \hline
  A             & 14.07 & 13.94 & 14.18 & 13.83 & 14.05 & 14.09 & 13.31  & 13.28  \\ %\hline
  B             & -0.48 & -0.92 & -1.74 & 0.91 &  -0.2447 & -2.86 &  0.26 & 1.21  \\ %\hline
  C              & -1.71 & -0.27 & 0.99 & -2.49 &  -1.4 & 3.18 &  -1.66 & -3.33 \\
  \hline

\end{tabular}  %\label{tabledrift}
\end{table*}
%%%%%%%%%%%%%%%%%%%%%%%%%%%%%%%%%%%%%%%%%%%%%%%%%%%%%%%%%%%%%%%%%%%%%%%%%%%%%%%%%%%%%%

\begin{figure*}[h]
\vspace*{1mm}
\begin{center}
\includegraphics[scale=1.0]{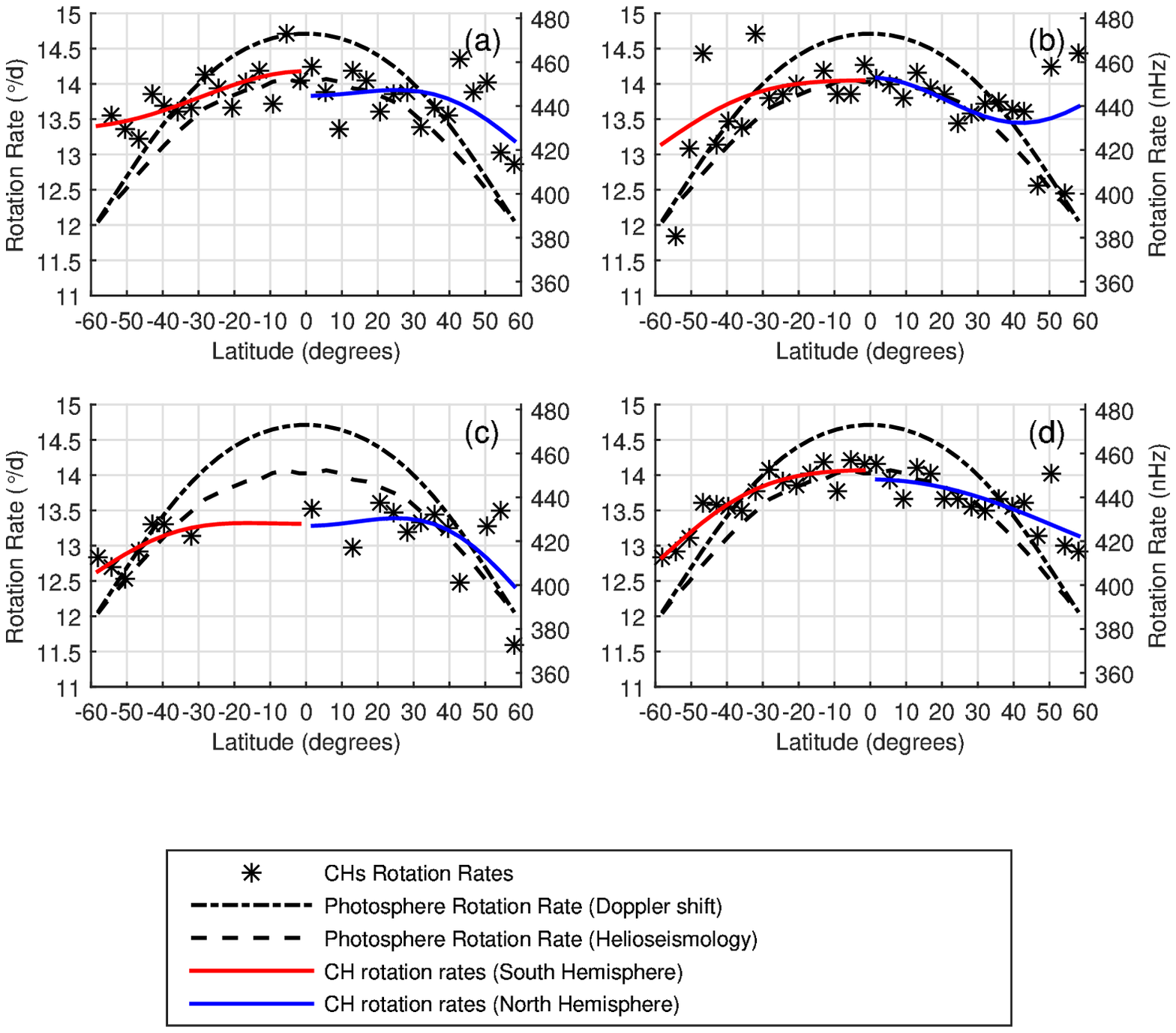}
\end{center}

\caption{CH sidereal rotation rates for 32 latitudinal zones $3.75^{\circ}$ wide, and their corresponding curves fitted for each hemisphere separately. The black dot-dashed line indicates the photosphere rotation rate obtained from Doppler shift measurements \citep{Snodgrass1990}. The black dashed line indicates the photosphere rotation rate obtained from the helioseismology \citep{Schou1998}. The black crosses are the rotation rates of CHs for the corresponding latitudes. The red curve corresponds to the rotation profile of CHs in the southern hemisphere, while the blue curve shows the rotation profile of CHs in the northern hemisphere.
Panel (a) provides information about the rotation rates of CHs with areas less than $10\;000\;Mm^{2}$; the corresponding error bar is 0.07 ${^\circ}/d$ (2.25 nHz); panel (b) shows results for CHs with areas from $10\;000$ to $40\;000\;Mm^{2}$, error=0.1 ${^\circ}/d$ (3.22 nHz);  panel (c) shows results for CHs with areas greater than $40\;000\;Mm^{2}$, error=0.09${^\circ}/d$ (2.89 nHz); and panel (d) shows the results for all CHs, error=0.07 ${^\circ}/d$  (2.25 nHz). Standard errors are calculated as $\sigma/\sqrt N$ for each area group.}
\label{Fig5}
\end{figure*}

After the calculation of the CH rotation rates, we concluded that their rotation profiles differ significantly from the photospheric differential rotation (Fig~\ref{Fig5}). These results indicate that the latitudinal gradients of the rotation rate are noticeably smaller than those observed in the photosphere (see Fig~\ref{Fig5}). The average rotation rate for all CHs is $13.86\;(\pm0.05)\;^{\circ}/d$. In the vicinity of the equator the velocity is approximately $14.15\;(\pm0.07)\;^{\circ}/d$ (i.e.\ a rotation period of 25.5 days). This value gradually decreases towards the poles, and in the vicinity of $\pm55-60^{\circ}$ latitude the rotation rates are $12.82\;(\pm0.07)\;^{\circ}/d$ in the southern hemisphere and $12.9\;(\pm0.07)\;^{\circ}/d$ in the northern hemisphere (Fig~\ref{Fig5}, panel (d)).

%%%%%%%%%%%%%%%%%%%%%%%%%%%%%%%%%%%%%%%%%%%%%%%%%%%%%%%%%%%%%%%%%%%%%%%%%%%%%%%%%%%%%%
%%%%%%%%%%%%%%%%%%%%%%%%%%%%%%%%%%%%%%%%%%%%%%%%%%%%%%%%%%%%%%%%%%%%%%%%%%%%%%%%%%%%%%

%%%%%%%%%%%%%%%%%%%%%%%%%%%%%%%%%%%%%%%%%%%%%%%%%%%%%%%%%%%%%%%%%%%%%%%%%%%%%%%%%%%%%%
\begin{figure*}[h]
\vspace*{1mm}
\begin{center}
\includegraphics[scale=1.0]{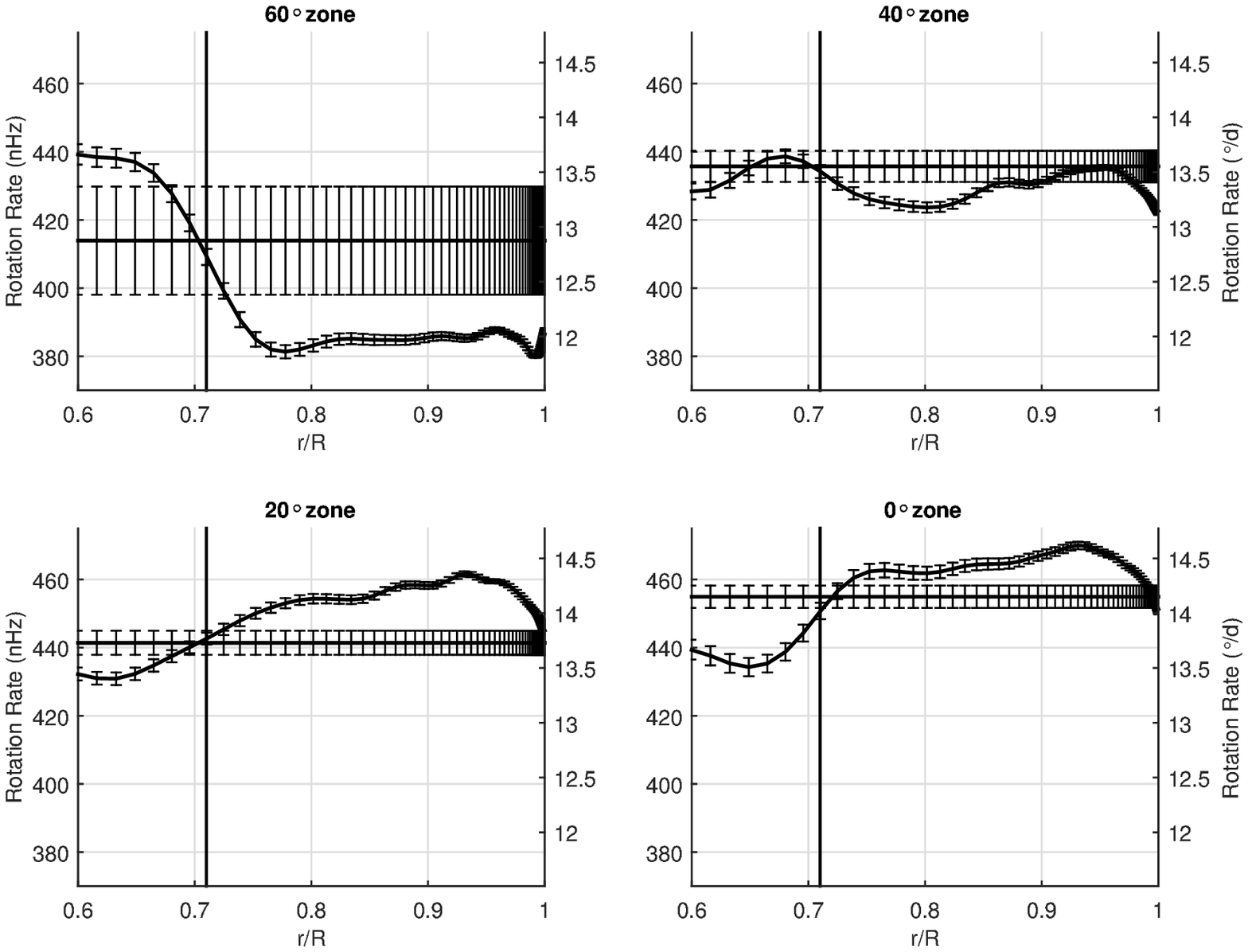}
\end{center}

\caption{Comparison between the CHs and solar interior rotation rates. The horizontal line represents the coronal hole sidereal rotation rate for the corresponding latitudinal zone. The curve represents the solar interior rotation profile for the same latitudinal zones. The crossing point indicates the depth where the CHs and the solar interior rotate with the same angular velocity. The vertical line corresponds to the bottom of the convective zone.}
\label{Fig6}
\end{figure*}
%%%%%%%%%%%%%%%%%%%%%%%%%%%%%%%%%%%%%%%%%%%%%%%%%%%%%%%%%%%%%%%%%%%%%%%%%%%%%%%%%%%%%%

The four panels in Fig~\ref{Fig5} present the rotation profiles for all, small-, medium-, and large-sized CHs, respectively. We have overlayed the photospheric rotation profiles estimated by two different methods: first,~from the helioseismology data \citep{Schou1998} (black dashed line) and, second,~the Doppler shift measurement result \citep{Snodgrass1990}(black dot-dashed line). The black asterisks denote the observed values of the rotation rates of CHs. The red and blue curves are obtained by fitting the parabolic law $A+Bsin^{2}(\theta)+Csin^{4}(\theta)$, representing the latitudinal profile of the rotation rate of CHs in the northern and southern hemispheres, respectively (Fig~\ref{Fig5}).

Each of the plots in Fig~\ref{Fig5} indicates that the rotation profiles of CHs differ from both estimates of the photospheric rotation. In particular,  the sidereal angular velocities of CHs, measured  with the low corona EUV data, are higher than those rates in the photosphere above 35-40 degrees latitude; we used a constant for transformation
from synodic to sidereal rotation rate given in \citet{Wittmann1996}. These results are in good agreement with other measurements of coronal rotation rates, for instance, those reported by  \citet{Mancuso2011}. The values of the coefficients that we obtained after the fitting are presented in Table~1.

%=====================
\section{Discussion}
%=====================
\subsection{Probable connection to the tachocline area}
%=====================

After several long-term investigations of solar photosphere rotation features, various authors have obtained different rotation profiles. We presented two illustrative examples for comparison in Fig~\ref{Fig5}. In fact, different observations can give slightly different profiles. However, it is most likely that all of these rotation profiles are located between the two curves obtained from observations by \citet{Snodgrass1990}, during the years  1967-1987, and from the helioseismological observation.

%%%%%%%%%%%%%%%%%%%%%%%%%%%%%%%%%%%%%%%%%%%%%%%%%%%%%%%%%%%%%%%%%%%%%%%%%%%%%%%%%%%%%%
\begin{figure*}[h]
\vspace*{1mm}
\begin{center}
\includegraphics[scale=1.0]{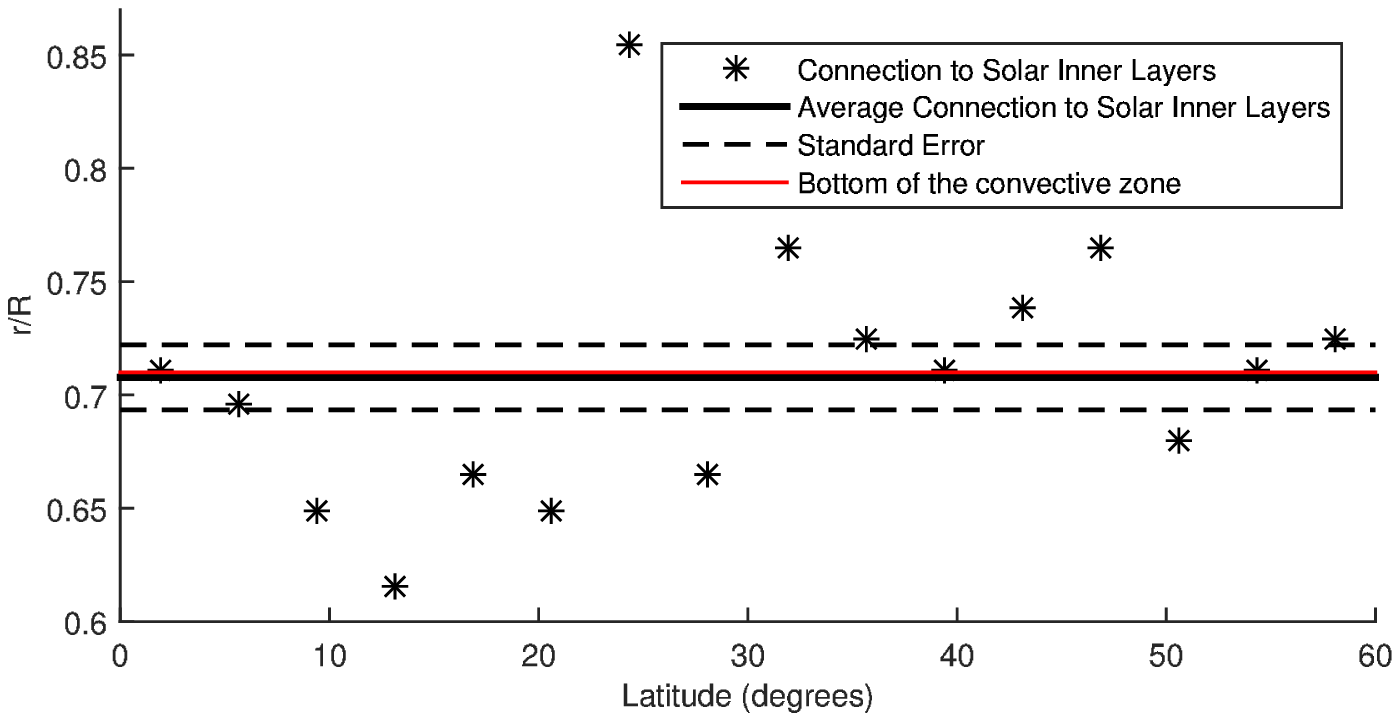}
\end{center}

\caption{Connection between the CHs and the solar interior.}
\label{Fig7}
\end{figure*}
%%%%%%%%%%%%%%%%%%%%%%%%%%%%%%%%%%%%%%%%%%%%%%%%%%%%%%%%%%%%%%%%%%%%%%%%%%%%%%%%%%%%%%

\begin{figure*}[h]
\vspace*{1mm}
\begin{center}
\includegraphics[scale=1.0]{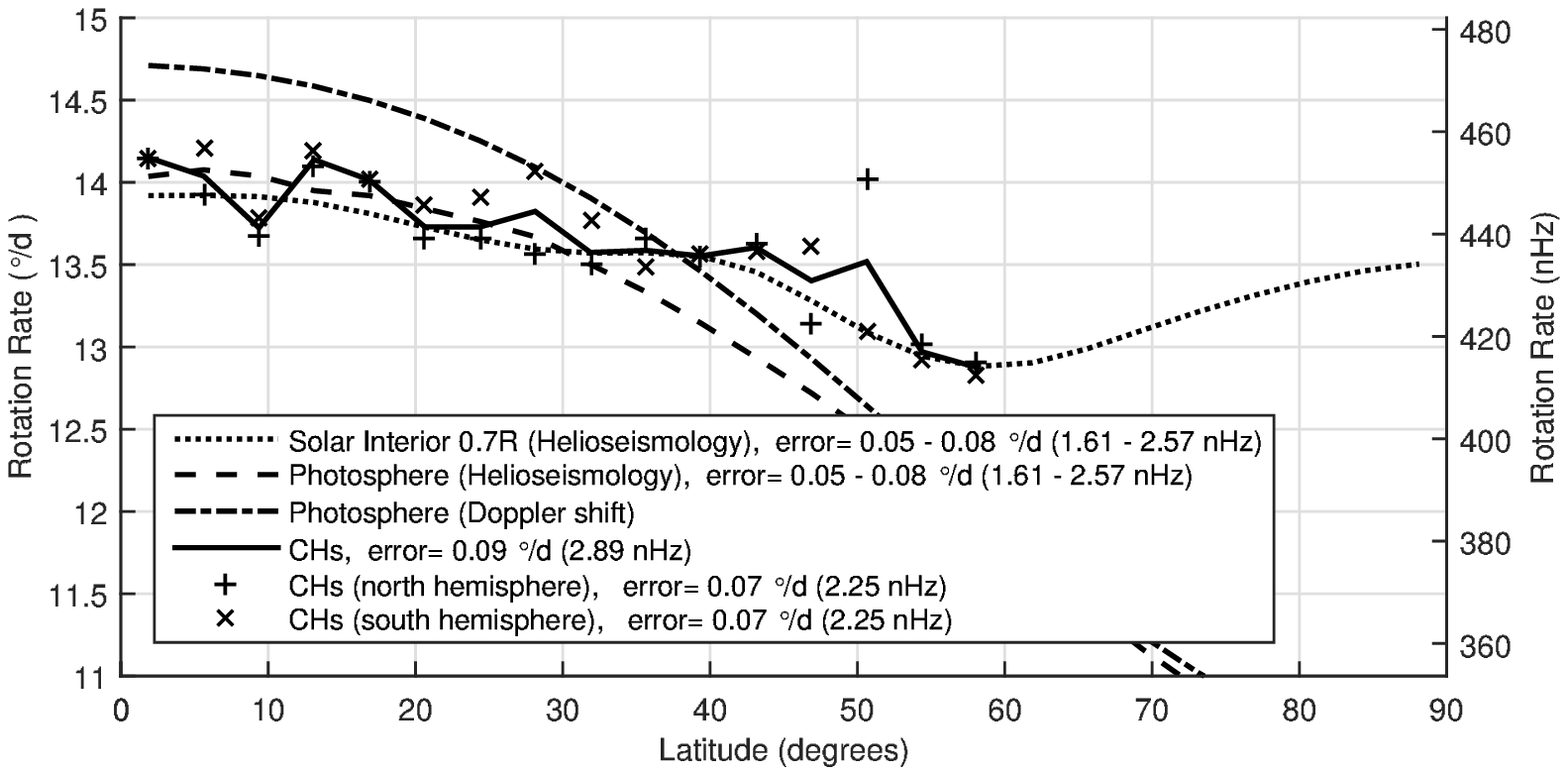}
\end{center}

\caption{Sidereal rotation profile comparison between different layers of the Sun. The dot-dashed line  indicates the photosphere rotation rate \citep{Snodgrass1990}; the dashed line indicates the photosphere rotation rate according to \citet{Schou1998}; the dotted line denotes the solar interior rotation rate at $0.71\;R_{\sun}$; the solid line is the rotation rate of the CHs; the plus signs are the rotation rates of the CHs in the northern hemisphere; and the cross signs are the rotation rates of the CHs in the southern hemisphere.}
\label{Fig8}
\end{figure*}

We compare the average angular velocities of CHs calculated in the 3.75-degree width latitudinal zones (in total 16 latitudinal zones) with the solar interior rotation rate values in the same latitudinal zones from 0.6 to 1$\;R_{\sun}$. For illustration, we presented in Fig~\ref{Fig6} only four examples for the latitudinal zones at 0$^{\circ}$, 20$^{\circ}$, 40$^{\circ}$, and 60$^{\circ}$. Now, in Fig~\ref{Fig7} we show which depth of the solar interior matches the angular velocity of CHs. The asterisks in Fig~\ref{Fig7} represent the region in which the solar interior rotates with the same rotation rate as the CHs in the corresponding latitudinal zone. These values vary from 0.62 to $0.85\;R_{\sun}$; while on average, it is $0.71\;(\pm0.014)\;R_{\sun}$. Fig~\ref{Fig8} represents two different photospheric rotation profiles (dot-dashed line and dashed line), the solar interior rotation profile at $0.71\;R_{\sun}$ (dotted line) and the angular velocity values of CHs combined for both hemispheres (solid line) and separately for the northern and southern hemispheres (plus and cross signs). The plots clearly show that the rotation rates of CHs directly follow the curve that corresponds to the solar internal rotation at the radius around $0.71\;R_{\sun}$. Moreover, in the vicinity of the equator the rotation rates of CHs are quantitatively similar to the photospheric rotation values obtained from the helioseismological observations; above 30$^{\circ}$ latitude the angular velocities of the photosphere decrease significantly, while the solar interior and CHs rotation profiles almost identically coincide with each other. Fig~\ref{Fig8} shows that the interior and CHs rotate faster than the photosphere above this latitude, and this is in accordance with other measurements \citep*{Mancuso2011}.

%%%%%%%%%%%%%%%%%%%%%%%%%%%%%%%%%%%%%%%%%%%%%%%%%%%%%%%%%%%%%%%%%%%%%%%%%%%%%%%%%%%%%%
%%%%%%%%%%%%%%%%%%%%%%%%%%%%%%%%%%%%%%%%%%%%%%%%%%%%%%%%%%%%%%%%%%%%%%%%%%%%%%%%%%%%%%

Our findings indicate that CHs could be linked to the deep layers of the Sun below the convective zone, in the tachocline, where the solar global magnetic field presumably originates. It turns out that although the rotation rates and photosphere angular velocities of CHs (helioseismological results) are quantitatively close to each other in the equatorial region, in general, the latitudinal characteristics of CHs rotation do not match any of the photospheric rotation profiles. These results may indicate that the magnetic field lines of the CHs are connected to the tachocline and the lower layers of the convection zone, approximately located at $0.71\;R_{\sun}$. We give a qualitative description of the possible physical mechanism that might provide link between CHs with deep subphotospheric layers in the conclusions.

Recently, similar results were obtained by \citet*{Hiremath13}, who investigated CHs rotation rates based on SOHO/EIT data during the years 2001-2008. Their average angular velocity ($438\;nHz$) corresponds to the solar interior rotation rate at $0.62\;(\pm0.10)$ solar radii.  This value is slightly smaller compared to the depth obtained by us, although the main conclusion that CHs magnetic field lines are connected to the inner layers of the Sun still remains.

%=====================
\section{Conclusions}
%=====================

Our goal was to evaluate the solar CH sidereal rotation rates and to study the differences in the rotation characteristics among CHs of different sizes. To achieve these objectives, we used the CH time series with data from the SPoCA, which uses the SDO/AIA data in the $193\;{\AA}$ wavelength. In total, we examined 540 objects in a 120X120-degree area on the solar disc from 1 January 2013  to 20 April 2015.

The analysis indicates that there was a north-south asymmetry in the distribution of CHs during the years 2013-2015. Approximately 60$\%$ of the detected objects were located in the northern hemisphere. In this epoch, CHs were presented at all latitudes. However, the largest and smallest objects were found only at high latitudes (cf.\ Fig~\ref{Fig2}).

The average angular velocity of the 540 investigated objects is $13.86\;(\pm0.05)\;^{\circ}/d$. In the equatorial area, the rotation rate is $14.15\;(\pm0.07)\;^{\circ}/d$ , while in the $\pm55-60^{\circ}$ latitudinal zone it decreases to approximately $12.9\;(\pm0.07)\;^{\circ}/d$. We found that the CH rotation rate is drastically different from the photospheric rotation rate.

The CH rotation profile almost perfectly coincides with the rotation profile of the tachocline and the lower layers of the convection zone at around $0.71\;R_{\sun}$. This fact gives an indication that CHs may be linked to the solar global magnetic field, which originates in the tachocline at this depth.

We examined objects only in the $\pm60^{\circ}$ latitudinal zone. Further study is needed to prove similar conclusions for latitudes above $\pm60^{\circ}$. Also, mathematical modelling, both analytical and numerical, is desirable to get further insight into the connection of the CH roots with the deep layers of the solar interior and global solar magnetic field.

In this work, we report the coincidence of the CH rotation rates with those in the lower convection zone, which allows us to propose the hypothesis that CHs may be linked to the deep solar layers through the configuration of the global magnetic field. Of course, a statement of such causality requires rigorous modelling of the mentioned field configurations and related connections. Pursuing of such modelling is not the primary aim of the current work and is somewhat away from its scope. However, we can make here some indicative remarks about a possible explanation of found coincidences. In a recent study,  \citet{Ruvzdjak2004} reported the similar analysis of rotation of sunspot groups with different age. The main conclusion of this work was that in the beginning their lifetime the young sunspot groups are plausibly anchored at higher depths (about $0.93\;R_{\sun}$), while the anchoring depth becomes more and deeper and may reach lower convection zone for those sunspot groups that
are recursive. This may be related to the gradual emergence of magnetic flux from tachocline and its connection to the magnetic features formed on the solar surface. On the other hand, there is the observation of the fact that CHs may originate from coronal funnels \citep{Tu2005} meaning that coronal holes and the fast wind patterns are tightly linked with the lower solar atmosphere and magnetic field topology there. Based on these reports, one can suppose that CHs may be formed and connected with the global residual magnetic field emerged in long-living recursive sunspot groups and thus provided a link between CHs and solar deeper layers. A recent work by \citet{mcintosh2014, mcintosh2015} has demonstrated that the sunspot cycle is driven by the rotation of the deep interior of the Sun. These authors have explained it in terms of the intra- and extra-hemispheric interaction between the overlapping activity bands of the 22-year magnetic polarity cycle, which are driven by the rotation of the deep interior of the Sun. At the same time there might be rather small CHs, which presumably belong to the group with an area less $10\;000\;Mm^{2}$, with short lifetimes (of the order of few days). These CHs can be linked with a residual magnetic field of weaker non-recursive sunspot groups, which are linked to the CHs anchored at higher layers. This intuitively drawn scenario is also in agreement with conclusions given by \citet{Nash1988}. Possibly this can be an indirect explanation of the presence of two intersection points between the helioseismic radial rotation profile and CH mean rotation rates. The proof or denying of this stated hypothesis needs a proper analytical and/or numerical modelling, which can become the subject of future studies.

\begin{acknowledgements}
The work was supported by European FP7-PEOPLE-2010-IRSES-269299 project - SOLSPANET. The work was also supported by Shota Rustaveli National Science Foundation grant DI-2016-52. Work of S.R.B. was supported under Shota Rustaveli National Science Foundation grants for doctoral students - PhDF2016\_204 and grant for young scientists for scientific research internships abroad IG/50/1/16. B.M.S. and M.L.K. acknowledge the support by the Austrian Fonds zur Foerderung der Wissenschaftlichen Forschung within the project P25640-N27 and Leverhulme Trust grant IN-2014-016. The work of T.Z. was supported by the Austrian Fonds zur Forderung der wissenschaftlichen Forschung under projects P25640-N27 and P26181-N27 and from the Georgian Shota Rustaveli National Science Foundation projects DI-2016-17 and 217146. A. K. was supported by NASA grant NNX14AB70G. M.L.K. additionally acknowledges the support of the FWF projects I2939-N27 and P25587-N27, and the grants No.16-52-14006, No.14-29-06036 of the Russian Fund for Basic Research. We are thankful to the anonymous referee for constructive comments that led to significant improvement of the manuscript content.

\end{acknowledgements}

\bibliography{mybib}

\end{document}